\documentclass[a4paper]{jpconf}

\usepackage{amsmath}
\usepackage{amsfonts}
\usepackage{amssymb}
\usepackage{dsfont}
\usepackage{graphicx}
\usepackage{slashed}
\usepackage{yfonts}
\usepackage{cite}
\usepackage{nicefrac}
\usepackage[usenames]{color}
\usepackage[colorlinks=true,linkcolor=blue,citecolor=blue,urlcolor=blue]{hyperref}

\definecolor{violet}{RGB}{111,0,255}
\definecolor{lred}{rgb}{1,0.90,0.7}

\newcommand{\ONE}{\mathds{1}} 

\begin{document}
\title{Hadronic Observables from Dyson--Schwinger and Bethe--Salpeter equations}

\author{\underline{H\`elios Sanchis-Alepuz} and Richard Williams}
\address{Institut f\"ur Theoretische Physik, Justus-Liebig--Universit\"at Giessen, 35392 Giessen, Germany}
\ead{helios.sanchis-alepuz@theo.physik.uni-giessen.de}
\ead{richard.williams@theo.physik.uni-giessen.de}

\begin{abstract}
In these proceedings we present a mini-review on the topic of the 
Dyson--Schwinger/Bethe--Salpeter approach to the study of relativistic 
bound-states in physics. In particular, we present a self-contained discussion 
of their derivation, as well as their truncation such that important symmetries 
are maintained.
\end{abstract}

\section{Introduction}
In Quantum Chromodynamics (QCD) the only observable objects are hadrons, 
which appear as bound-states of the elementary (but not observed) quark and 
gluon degrees of freedom. Consequently, most phenomenological aspects of QCD are 
essentially non-perturbative problems and require an appropriate framework for 
their study. One such approach is the combination of Dyson--Schwinger (DSE) and 
Bethe-Salpeter (BSE) equations that provide a means to study the 
non-perturbative properties of hadrons -- at the microscopic level -- without 
abandoning \emph{a priori} the principles of QCD as a scale dependent continuum 
quantum field theory. We devote these proceedings to a concise exposition of 
known results about how 
this framework can be systematically constructed as well as a description of the 
main technical issues faced upon solving the DSEs and BSEs in combination.

We refrain, for reasons of space, from including here results for observable 
data that can already be found in the literature~\cite{Bashir:2012fs,Eichmann:2013afa,Cloet:2013jya}. 
Meson spectra have been thoroughly studied in~\cite{Jain:1993qh,Roberts:1996jx,Burden:1996nh,Roberts:1997vs,Maris:1997hd,Maris:1997tm,Maris:1999nt,Bender:2002as,Maris:2003vk,Holl:2004fr,Holl:2004un,Maris:2006ea,Cloet:2007pi,Fischer:2009jm,Krassnigg:2009zh,Krassnigg:2010mh,Blank:2011ha,Qin:2011dd,Qin:2011xq,Chang:2011vu,Fischer:2014xha,Hilger:2014nma,Gomez-Rocha:2014vsa,Fischer:2014cfa,Hilger:2015hka} 
and references therein. Baryons have been studied in the quark-diquark 
approximation~\cite{Oettel:1998bk,Oettel:2000jj,Oettel:2002wf,Nicmorus:2010sd,Chen:2012qr,Segovia:2013rca,Segovia:2013uga,Segovia:2014aza,Pitschmann:2014jxa}, and also as three-body 
objects~\cite{Eichmann:2009qa,Eichmann:2011ej,Sanchis-Alepuz:2013iia,Sanchis-Alepuz:2014mca,SanchisAlepuz:2011jn,Sanchis-Alepuz:2014sca}.  Other observables of enormous experimental interest 
in deep inelastic scattering such as PDAs, PDFs, GPDs are also starting to be 
investigated~\cite{Chang:2014lva,Shi:2014uwa,Gao:2014bca,Chang:2013nia,Mezrag:2014jka}. 
Most of these calculations were performed upon truncating the (anti)quark-quark 
interaction to a single gluon exchange, the so-called Rainbow-Ladder truncation. 
This can be seen as the leading order interaction mechanism in the systematic 
expansion to be described below. While it provides accurate results for 
ground-state pseudoscalar and vector mesons, and also for ground state baryons, 
it shows clear deficiencies in all other channels. For this reason, current 
interest lies is the inclusion of interaction mechanisms beyond the leading 
one~\cite{Bender:1996bb,Watson:2004kd,Bhagwat:2004hn,Matevosyan:2006bk,Fischer:2007ze,Fischer:2008sp,Fischer:2008wy,Fischer:2009jm,Williams:2009wx,Chang:2009zb,Heupel:2014ina,Sanchis-Alepuz:2014wea,Aguilar:2014lha,Mitter:2014wpa,Braun:2014ata,Sanchis-Alepuz:2015qra}, and in the
extension to glueball~\cite{Kellermann:2012fla,Meyers:2012ka,Sanchis-Alepuz:2015hma} and tetraquark bound-states~\cite{Heupel:2012ua}. 

The main purpose, therefore, of this contribution is to make it apparent that one 
is not pursuing a blind hunt for missing interaction mechanisms, but that there 
is a step-by-step programme for their inclusion in a systematic manner.

\section{Dyson--Schwinger and Bethe--Salpeter equations}
A complete description of a continuum quantum field theory, and in particular of 
QCD, is given when all the (infinitely many) Green's functions of the theory 
are known. Functional methods deal with these without abandoning the realm of 
continuum physics, thus providing an approach complementary to that of lattice
calculations. 
In particular, Dyson-Schwinger equations
constitute an infinite set of coupled, non-linear integral equations for the 
full Green's functions of the theory. We define in this section the basic 
concepts to be used later on.

\subsection{Functional Methods}
The generating functional for full Green's functions corresponding to a 
classical Euclidean action $\textrm{S}[\phi]$ is given by
\begin{equation}
	Z[J] = \int \mathcal{D}\phi e^{-i\left( \textrm{S}[\phi] + J_i 
\phi_i\right)}
	     = \left< \exp \left( -i J_i \phi_i \right)\right> \;,
\end{equation}
where $J_i\phi_i$ denotes $\int d^4x J_a(x) \phi_a(x)$
with $i$ a superindex that subsumes possible discrete (collectively denoted $a$) 
and continuous indices $x$. This functional is normalised such that $Z[0]=1$.
The $\left< \cdot \right>$ indicate the weighted 
functional average with sources subsequently set to zero. In the presence of 
external sources one writes $\left< \cdot \right>_J$.

The generating functional for connected Green's functions is given by
\begin{equation}
	W[J] = -i\ln Z[J]\;,
\end{equation}
which can be shown by a Taylor expansion. Furthermore, the generating functional 
for one-particle irreducible vertex functions (the effective action) is obtained 
by a Legendre transform
\begin{equation}
	\Gamma[\phi^c] = W[J] - \phi_i^c J_i\;,\qquad\textrm{with}\,\, \phi_i^c 
	= \left<\phi_i\right>_J = \frac{\delta W[J]}{\delta J_i}\;.
\end{equation}

\subsection{Dyson--Schwinger Equation}
The observation that the integral of a total derivative vanishes, true so long 
as the functional measure is invariant under spacetime translations of the 
field variables
\begin{equation}
	\left< \frac{\delta}{\delta \phi_i} S[\phi] - J_i\right>_J = 0\;,
\end{equation}
can be used to obtain the Dyson--Schwinger equations (DSEs). 
Equivalently, one can view the DSEs as a 
consequence of the Ward-Identity associated with translational invariance. These
DSEs are the quantum field theoretic equivalent of the classical equations of 
motion. Upon a vertex expansion we generate the infinite tower of non-linear 
integral equations that relate the fundamental Green's functions of the quantum 
field theory to one-another. In a very dense notation~\cite{Alkofer:2000wg} the 
DSEs for proper $n$-point Green's functions may be obtained from
\begin{equation}
	\frac{\delta\Gamma[\phi]}{\delta \phi_i} - \frac{\delta S}{\delta\phi_i}\left[\phi 
	+ \frac{\delta^2W}{\delta J\delta J}\frac{\delta}{\delta\phi}\right] = 0\;,
\end{equation}
by taking vacuum expectation values $\left< \cdot \right>$ of the $n$th 
derivative of the functional.

\subsection{Bound state equations}
Consider an $n$-particle Green's 
function $G^{(n)}(p_1,\ldots,p_n)$ describing the evolution of an $n$-particle 
system (each carrying momentum $p_i$), and its amputated counterpart the scattering scattering matrix 
$T^{(n)}(p_1,\ldots,p_n)$
\begin{align}\label{eq:scattering_matrix}
    G^{(n)} = G^{(n)}_0+G^{(n)}_0T^{(n)}G^{(n)}_0\,,
\end{align}
with $G_0^{(n)}$ the disconnected product of $n$-propagators. Then we may obtain 
$T^{(n)}$ from the Dyson equation
\begin{align}\label{eq:scattering_matrix_DE}
    T^{(n)}=K+KG^{(n)}_0T^{(n)}\,,
\end{align}
with $K$ the $2,3,$-, \ldots, $n$-particle irreducible interaction kernel. 
When the system forms a bound state, the momentum-space function $T^{(n)}$ 
develops a pole at 
$P^2=-M^2$ (in Euclidean spacetime) with $P^\mu=\sum_i^n p^\mu_i$ the total momentum. At the pole we define
\begin{align}\label{eq:def_BSEamplitudes}
	T^{(n)}\sim\mathcal{N}\frac{\Psi\bar{\Psi}}{P^2+M^2}\,,
\end{align}
where $\mathcal{N}$ is a state-dependent normalisation factor, $\Psi$ is the 
bound-state Bethe-Salpeter amplitude and $\bar{\Psi}$ its charge conjugate. 
Inserting this ansatz into the Dyson equation for the $T$-matrix and equating 
residues yields the homogeneous Bethe-Salpeter equation
\begin{align}\label{eq:compactBSE}
	\Psi=KG^{(n)}_0\Psi \,,
\end{align}
or for its conjugate
\begin{align}\label{eq:compactBSE-conj}
 	\bar{\Psi}K^{-1}=\bar{\Psi}G^{(n)}_0 \,.
\end{align}
It can also be shown that the correct normalisation condition for 
Bethe-Salpeter amplitudes is \cite{Cutkosky:1964zz,Nakanishi:1965zza}
\begin{equation}
\bar{\Psi}\left(\frac{dG^{(n)}_0}{dP^2}-i\frac{dK^{-1}}{dP^2}\right)\Psi=1.
\end{equation}

Finally, Bethe-Salpeter amplitudes are the amputated versions of the 
Bethe-Salpeter 
wave functions $\varphi=G^{(n)}_0\Psi$. Using this notation, we can rewrite 
\eqref{eq:compactBSE} as
\begin{align}\label{eq:compactBSE_with waves}
	\Psi=K\varphi \,,
\end{align}
The interaction kernels $K$ contain a sum of 
infinitely many terms. If they are known, one can then solve the BSEs and obtain 
all the information about 
the bound states of the theory.

\section{Deriving the Bethe-Salpeter kernel}
The effective action provides a systematic means to derive (generalised) 
Bethe-Salpeter kernels. This not only enables new interaction mechanisms to be 
included in a controlled way, but ensures that relevant symmetries are 
maintained even when approximations are made. Here, we outline the main ideas 
behind this procedure for the case of two- and three-body bound-state equations 
following Refs.~\cite{Fukuda:1987su,Komachiya:1989kc,McKay:1989rk}.

\subsection{Local case}
We start with the case of local fields, introducing the main ideas applicable to 
the bi- and tri-local fields necessary for the study of bound-state equations. 
First, let us consider the generating functional for a purely fermionic action 
and introduce source terms $J_k(x)$ for a local operator 
$\mathcal{O}_k(x)=\left\{\psi(x),\bar{\psi}(x),\psi(x)^2,\ldots\right\}$, with 
$k$ an index describing all discrete indices of the operator and, also, 
distinguishing the different possible local operators
\begin{align}
	Z[J] &= \mathcal{N} \int \mathcal{D}\psi e^{-i\left(\mathcal{S}[\psi]
                 +\int J_k(x) \mathcal{O}_k(x)\right)}\,,\label{eq:partition_function}
\end{align}
where $\mathcal{N} = 1/Z[J]\big|_{J=0}$. Defining, as above, the generating 
functional for 
connected Green's functions as $W[J]=-i \log Z[J]$, the one-particle irreducible 
(1PI) effective action is given by the Legendre transformation
\begin{align}\label{eq:EffAction1PI}
	\Gamma[\Psi]=W[J]-\int d^4x J_k(x)\Psi_k(x)\,,
\end{align}
with the \textit{classical} (super)field defined as $\Psi_k(x) = \delta W[J]/\delta J_k(x)$. It can be determined in the absence of sources through satisfaction of the stationary condition
\begin{align}\label{eq:EqMot1PI}
    \frac{\delta\Gamma[\Psi]}{\delta\Psi_k}=0\,.
\end{align}
Consider a solution of this stationary condition, $\Psi^{(0)}_k(x)$. We call it 
\emph{stable} if there exists solution $\Psi^{(1)}_k(x)=\Psi^{(0)}_k(x)+\Delta\Psi_k(x)$ 
with $\Delta\Psi_k(x)$ \emph{suitably infinitesimal}. Expanding around 
$\Psi^{(0)}_k(x)$ yields
\begin{align}
    \frac{\delta\Gamma[\Psi]}{\delta\Psi_k}\bigg|_{\Psi_k=\Psi^{(1)}_k}=\frac{\delta\Gamma[\Psi]}{\delta\Psi_k}\bigg|_{\Psi_k=\Psi^{(0)}_k}-\int d^4x'\frac{\delta^2\Gamma[\Psi]}{\delta\Psi_k(x)\delta\Psi_\ell(x')}\bigg|_{\Psi_k=\Psi^{(0)}_k}\Delta\Psi_\ell(x')+\dots=0~,
\end{align}
with a minus sign from exchanging the order of derivatives. To first order this 
equation implies
\begin{align}
    \int d^4x'\frac{\delta^2\Gamma[\Psi]}{\delta\Psi_k(x)\delta\Psi_\ell(x')}\bigg|_{\Psi_k=\Psi^{(0)}_k}\Delta\Psi_\ell(x')=0~.
\end{align}
That is, the solution $\Psi^{(0)}_k(x)$ is stable if $\Delta\Psi_k(x)$ is an 
eigenvector of $\delta^2\Gamma / \delta\Psi\delta\Psi$ with zero eigenvalue. To 
interpret this, let us recall the following identity
\begin{align}
    \int d^4x''\frac{\delta^2\Gamma[\Psi]}{\delta\Psi_\ell(x)\delta\Psi_k(x'')}\frac{\delta^2 W[J]}{\delta J_k(x'')\delta J_m(x')}=-\delta_{\ell m}\delta^{(4)}(x-x')~.\label{eq:Id_W_G}
\end{align}
Now, consider an eigenvector $\xi_k(x)$ of $\delta^2 W / \delta J\delta J$ with 
eigenvalue $1/\lambda$
\begin{align}
    \int d^4x'\frac{\delta^2 W[J]}{\delta J_k(x)\delta J_\ell(x')}\xi_\ell(x')=\frac{1}{\lambda}\xi_k(x)~.
\end{align}
Using the identity \eqref{eq:Id_W_G} we arrive at
\begin{align}
    \int d^4x'\frac{\delta^2 \Gamma[\Psi]}{\delta \Psi_k(x)\delta \Psi_\ell(x')}\xi_\ell(x')=\lambda\xi_k(x)~.
\end{align}
Therefore, $\xi_k(x)$ is also an eigenvector of $\delta^2\Gamma / \delta\Psi\delta\Psi$ 
with eigenvalue $\lambda$. In particular, the perturbation $\Delta\Psi_k(x)$ 
corresponds to an eigenvector with $\lambda=0$, wich in turn is related to a 
pole in $\delta^2 W / \delta J\delta J$. In summary, stable solutions for 
$\Psi_k(x)$ are associated to poles of the connected Green's functions given by 
$\delta^2 W / \delta J\delta J$ (e.g. propagators if $\mathcal{O}(x)=\psi(x)$).

\subsection{Bilocal case: Two-body bound-states}
For the study of two body problems we add a bilocal term 
$K^k_{rs}(x,y)\mathcal{O}^k_{rs}(x,y)$ to the partition 
function~\eqref{eq:partition_function}, with $k$ again distinguishing the 
different bilocal operators and the other discrete indices described by $r,s$. 
Specialising to the case of fermion-antifermion bound states (mesons) we limit 
the discussion to $\mathcal{O}_{rs}(x,y)=\psi_r(x)\bar{\psi}_s(y)$ and 
antisymmetric sources $K_{rs}(x,y)=-K_{sr}(y,x)$. Then, in addition to the 
classical field $\Psi_k(x)$ we have the connected two-point Green's function
\begin{align}\label{eq:SemiClassicalPropagator}
	\frac{\delta W[J,K]}{\delta K_{ab}(x,y)}=\frac{1}{2}\left(\Psi_a(x)\bar{\Psi}_b(y)+G_{ab}(x,y)\right)\,.
\end{align}
The 2PI effective action is then defined as
\begin{align}
	\Gamma[\Psi,G]=W[J,K]-\int d^4x J_k(x)\Psi_k(x)&-\frac{1}{2}\int d^4xd^4y K_{ab}(x,y)\Psi_a(x)\bar{\Psi}_b(y)\\
                                               &-\frac{1}{2}\int d^4xd^4y K_{ab}(x,y)G_{ab}(x,y)~,\label{eq:2PILegTr}
\end{align}
which admits, in addition to \eqref{eq:EqMot1PI} and in the absence of sources, 
the stationary condition 
\begin{align}
    \frac{\delta\Gamma[\Psi,G]}{\delta G_{ab}(x,y)}=0\,.\label{eq:EqMot2PI}
\end{align}
This is equivalent to the Dyson--Schwinger equation for the fermion propagator, 
as we show below.

Following the ideas of previous section, a solution $G^{(0)}_{ab}(x,y)$ of 
\eqref{eq:EqMot2PI} is stable if there exists a perturbed solution 
$G^{(1)}_{ab}(x,y)=G^{(0)}_{ab}(x,y)+\Delta G_{ab}(x,y)$ determined by a 
non-trivial solution of
\begin{align}
    \int d^4x'd^4y'\frac{\delta^2\Gamma[\Psi,G]}{\delta G_{ab}(x,y)\delta G_{a'b'}(x',y')}\bigg|_{G=G^{(0)}}\Delta G_{a'b'}(x',y')=0~.\label{eq:BSE_obscure}
\end{align}
We show below that this is indeed equivalent to the usual Bethe-Salpeter 
equation for a fermion-antifermion bound state. Nevertheless this can be seen 
here, analogous to the local case, through the fact that a solution of 
\eqref{eq:BSE_obscure} is related to a pole in $\delta^2 W / \delta K\delta K$, 
\emph{viz.} to a pole in a four-point Green's function.

\subsubsection{Relation to the Bethe-Salpeter equation}
It is known~\cite{Cornwall:1974vz} that the 2PI effective action can be written 
as
\begin{align}
    \Gamma[\Psi,G]=\textrm{S}[\Psi]+iTr\log G-iTr G_0^{-1}G+\Gamma_2[\Psi,G]~,
\end{align}
where $\Gamma_2[\Psi,G]$ contains two-particle irreducible diagrams only and 
$G_0$ is the classical propagator. The stationary condition for $G$ gives, as 
promised above, the Dyson-Schwinger equation for the propagator. Indeed, taking 
a functional derivative with respect to the propagator $G$
\begin{align}
    \frac{\delta\Gamma[\Psi,G]}{\delta G_{ab}(x,y)}=iG^{-1}_{ab}(x,y)-iG_{0,ab}^{-1}(x,y)+\frac{\delta\Gamma_2[\Psi,G]}{\delta G_{ab}(x,y)}=0~,
\end{align}
where we used $\delta Tr\log G=G^{-1}\delta G$. Defining the self-energy as 
$\Sigma=-i\delta \Gamma_2 / \delta G$ we can rewrite the stationary condition 
for $G$ as
\begin{align}\label{eq:fermionselfenergy}
	G^{-1}_{ab}(x,y)=G_{0,ab}^{-1}(x,y)-\Sigma_{ab}(x,y)\;,
\end{align}
which is the gap equation for the fermion propagator. If $G^{(0)}$ is one of the 
solutions of the gap equation, we can take one further functional derivative 
with respect to the propagator $G$ and rewrite \eqref{eq:BSE_obscure} as
\begin{align}
    0 &= \int d^4x'd^4y'\frac{\delta^2\Gamma[\Psi,G]}{\delta G_{ab}(x,y)\delta G_{a'b'}(x',y')}\bigg|_{G=G^0}\Delta G_{a'b'}(x',y')\nonumber \\
      &=\int d^4x'd^4y' \left(-G^{(0)}_{ac}(x,x')G^{(0)}_{db}(y,y')+K_{ab;cd}(x,y;x',y')\right)\Delta G_{cd}(x',y')\,,\label{eq:BSE_clear}
\end{align}
where we used $\delta M^{-1}_{ij} / \delta{M_{kl}}=-M^{-1}_{ik}M^{-1}_{lj}$. 
This is precisely the Bethe-Salpeter equation \eqref{eq:compactBSE_with waves} 
for a Bethe-Salpeter wave function 
$\Delta G$, with the quark-antiquark interaction kernel given by
\begin{align}\label{eq:bsekernel}
    K_{ab;cd}(x,y;x',y')=-\frac{\delta\Sigma_{ab}(x,y)}{\delta G_{cd}(x',y')}\bigg|_{G=G^{(0)}}~.
\end{align}
One then sees immediately that the interaction kernel is obtained by 
functionally \textit{cutting} propagator lines from the self-energy. Note that 
the solution $G=G^{(0)}$ is inserted into the self-energy only after the cutting 
has been performed.

\subsubsection{Goldstone bosons. Chiral symmetry.}
We show here that when the meson BSE is derived using the methods outlined 
above, the spontaneous breaking of chiral symmetry is accompanied by the 
appearance of a pseudoscalar massless bound  state (the Goldstone boson).
More details can be found in~\cite{Munczek:1994zz}. Consider that under a global 
SU(2) chiral transformation
\begin{align}
    \Psi'=e^{i\gamma_5\tau^a\theta}\Psi\qquad~,\qquad 
G'=e^{i\gamma_5\tau^a\theta}G e^{i\gamma_5\tau^{\dag a}\theta}~,
\end{align}
the 2PI effective action (we ignore the dependence on the fields, as 
it plays no role in this discussion) is invariant $\Gamma[G']=\Gamma[G]$. One 
then has, for infinitesimal transformations
\begin{align}
\delta\Gamma[B]=\frac{\theta\delta\Gamma}{\delta 
G_{a'b'}\left(x',y'\right)}\{i\gamma_5\tau^a,G\left(x',y'\right)\}_{a'b'}=0~.
\end{align}
Taking one further derivative with respect to $G$ yields
\begin{align}
\frac{\delta^2\Gamma}{\delta 
G_{ab}\left(x,y\right)G_{a'b'}\left(x',y'\right)}\{i\gamma_5\tau^a,G\left(x'
,y'\right)\}_{a'b'}+\frac{\delta\Gamma}{\delta 
G_{ac}\left(x,y\right)}\tau^a\gamma_{5,cb}+\tau^a\gamma_{5,ac}\frac{\delta\Gamma
}{\delta 
G_{cb}\left(x,y\right)}=0~.
\end{align}
When we set $G$ to the solution of the stationary condition, $G=G^{(0)}$, the 
equation simplifies
\begin{align}
\frac{\delta^2\Gamma}{\delta 
G_{ab}\left(x,y\right)G_{a'b'}\left(x',y'\right)}\bigg|_{G=G^{(0)}}\{
\gamma_5\tau^a,G^{(0)}\left(x'
,y'\right)\}_{a'b'}=0~.
\end{align}
If chiral symmetry is spontaneously broken then $\{
\gamma_5\tau^a,G^{(0)}\left(x
,y\right)\}$ is non-vanishing, which means that there exists a solution of the 
BSE with pseudoscalar quantum numbers. Moreover, since 
$G^{(0)}(x,y)=G^{(0)}(x-y)$, the solution corresponds to one that has vanishing 
total momentum after transforming to momentum space.

The key observation here is that, even after the effective action is truncated 
to some loop order, as long as the truncated action is invariant and both the 
quark self-energy and the meson BSE kernel are derived by taking functional 
derivatives of it, Goldstone's theorem will hold without fine tuning.
We finally note that the arguments given here translate identically to any 
global symmetry of the effective action.

\subsection{Tri-local: Three-body bound-states}
The extension of the above formulae to the three-body bound state case is rather 
straightforward \cite{Komachiya:1989kc} if one adds to the partition function a 
source term for the trilocal operators of interest, 
$R^k_{rst}(x,y,z)V^k_{rst}(x,y,z)$. After a Legendre transformation, the 
effective action acquires an explicit dependence on the three-body vertex $V$. 
Note that depending on whether source terms for the proper tri-vertices of the 
theory are added or not, one is dealing with the 2PI or the 3PI effective 
action, supplemented with an extra vertex for the three-body bound state.

Focusing here on the case of a baryonic bound state, we introduce sources $R$ 
and $\bar{R}$ for the operators $V=\psi_r(x)\psi_s(y)\psi_t(z)$ and 
$\overline{V}=\bar{\psi}_r(x)\bar{\psi}_s(y)\bar{\psi}_t(z)$, respectively. 
Using the stability arguments laid above for a would-be solution $V^{(0)}$ of 
the 
stationary condition $\delta \Gamma / \delta V =0$ then lead to the following 
three-body bound state equation (see \cite{Komachiya:1989kc} for a detailed 
derivation)
\begin{align}
    \int d^4x'd^4y'd^4z'\frac{\delta^2\Gamma[\Psi,G,V]}{\delta \overline{V}_{rst}(x,y,z)\delta V_{r's't'}(x',y',z')}\bigg|_{G=G^{(0)},V=V^{(0)}}\Delta V_{r's't'}(x',y',z')=0\,.\label{eq:3bBSE}
\end{align}
Note that only mixed derivatives with respect to $V$ and $\overline{V}$ do not 
vanish identically when one sets $V=V^{(0)}$ and 
$\overline{V}=\overline{V}^{(0)}$.

It is important to comment here that for the important cases of mixed 
quark-flavour states, and indeed mixed states in general, the procedure just 
outlined proceeds identically, only with the introduction of mixed propagators
and vertices. 
Generally speaking, one could say that for each bound state of interest, one 
inserts the appropriate vertex in the effective action, takes 
functional derivatives with respect to it, and sets it to its vacuum value at 
the end. The fact that for flavour-diagonal mesons this vertex coincides with 
the quark propagator is therefore merely accidental.

\section{Internal structure: form factors}
The prototypical experiment for probing the internal structure of hadrons 
consists of (in)elastic scattering of a particle on a hadron. Such interactions 
are mediated by gauge fields, thus the theoretical study of hadrons must couple 
gauge fields so that relevant symmetries are preserved, as well as maintaining 
internal consistency of the theoretical framework. At the level of Green's 
functions, a method coined \textit{gauging of equations} exists that ensures 
current conservation~\cite{Haberzettl:1997jg,Kvinikhidze:1998xn,Kvinikhidze:1999xp}.

The method of \textit{gauging} can be best understood through the introduction 
of a new source term in the partition function, $-\mathcal{J}^\mu (x)A_\mu(x)$, 
where $A^\mu$ is the external, non-dynamical gauge field and $\mathcal{J}^\mu$ 
is the current that couples to it. For example, consider the $2n$-point 
fermionic Green's function
\begin{align} 
	G^{(2n)}(x_1\dots x_n;x'_1\dots x'_n)=
	\langle 0| T\left[\psi^{1}(x_1)\dots\psi^{n}(x_n)\bar{\psi}^{1}(x'_1)\dots\bar{\psi}^{n}(x'_n)\right] |0\rangle\,.
\end{align}
Then, its coupling to an external gauge field $A_\mu$ would be
\begin{align}
 G^{(2n),\mu}(x_1\dots x_n;x'_1\dots x'_n;y)=
\langle 0| T\left[\psi^{1}(x_1)\dots\psi^{n}(x_n)\bar{\psi}^{1}(x'_1)\dots\bar{\psi}^{n}(x'_n)\mathcal{J}^\mu(y)\right] |0\rangle\,,
\end{align}
which is called the \emph{gauged Green's function}. It can be obtained by 
applying a functional derivative with respect to $A^\mu$ to the original Green's 
function (we drop from now on the superindices indicating the number of 
particles)
\begin{align}
 G^\mu=-\left.\frac{\delta}{\delta A_\mu}G\right\lvert_{A=0}\,,
\end{align}
and setting the external field to zero.
It hence follows the ordinary rules for derivatives when acting, for instance, 
on products of functions.
The proper $n$-body vertex, $J^\mu$, for the coupling of the n-body system to 
the field $A_\mu$ is defined as
\begin{align}
 	G^\mu=G_0J^\mu G_0~.
\end{align}
with $G_0$ again the product of full propagators. In particular, for $n=1$ this 
defines the proper (fermion-photon) vertex 
$\Gamma^\mu$ 
\begin{align}\label{eq:definition_qphvertex}
 	S^\mu=S\Gamma^\mu S~,
\end{align}
as a result of gauging the full propagator $G^{(2)}=S$.
A useful relation follows from gauging $(SS^{-1})^\mu$ and using the identity 
$\mathds{1}^\mu=0$
\begin{align}\label{eq:definition_qphvertex_inverse}
 	(S^{-1})^\mu=-S^{-1}S^\mu S^{-1}=-\Gamma^\mu~.
\end{align}

For the following it is in general more convenient to work with the 
amputated Green's function, \textit{i.e.} the scattering matrix $T$ defined 
in~\eqref{eq:scattering_matrix}. Consider then a hadron described by the 
following Dyson equation
\begin{align}
 	T=K+K G_0 T~,
\end{align}
with $K$ the interaction 
kernels derived using the prescriptions given above. We can gauge this equation, 
following the rules of differentiation, to obtain
\begin{align}
 	T^\mu=K^\mu+K^\mu G_0T+KG_0^\mu T+K G_0T^\mu~,
\end{align}
which can be rewritten using \eqref{eq:scattering_matrix_DE} as
\begin{align}\label{eq:gaugedT}
 	T^\mu&=\left(\mathds{1}-KG_0\right)^{-1}\left(K^\mu+K^\mu G_0T+ KG_0^\mu T\right) \nonumber\\
    &=T\left(K^{-1}K^\mu K^{-1}+G_0^\mu\right)T~.
\end{align}

At the bound state poles, one can introduce a bound-state electromagnetic 
current $J^\mu$ in a similar fashion as in \eqref{eq:def_BSEamplitudes}
\begin{align}
 	T^{\mu}\sim\frac{\Psi_f}{P_f^2+M_f^2}J^\mu\frac{\bar{\Psi}_i}{P_i^2+M_i^2}~,
\end{align}
where $M_{i,f}$ and $\Psi_{i,f}$ are the initial and final bound-state masses 
and amplitudes, respectively. From \eqref{eq:gaugedT}, \eqref{eq:compactBSE} and 
\eqref{eq:compactBSE-conj}, we arrive at
\begin{align}\label{eq:FFeq_compact}
 	J^\mu=\bar{\Psi}_f \left(G_0^\mu+G_0 K^\mu G_0\right)\Psi_i~.
\end{align}

\begin{figure}[!t]
\begin{center}
\includegraphics[scale=1.5]{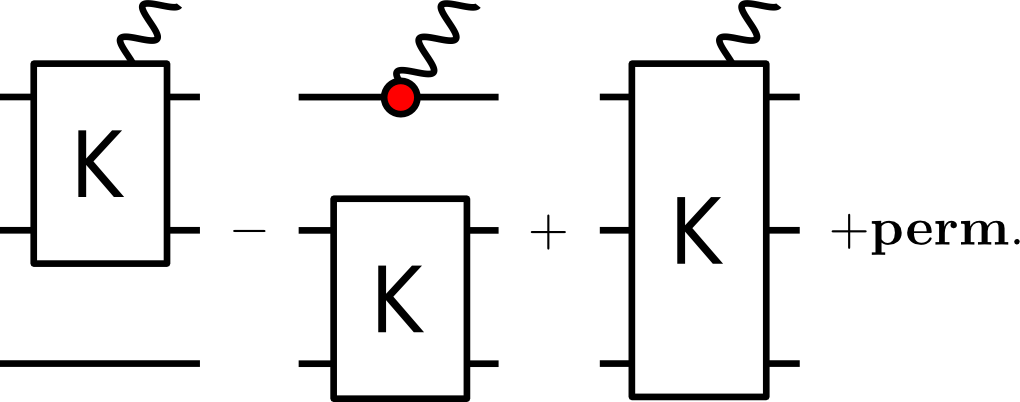}
\caption{\label{fig:FFeq_full}Gauging of the three-body kernel, $K^\mu$, as 
illustrated in \eqref{eq:gauged_kernel}. Note the minus sign which prevents the 
over counting of diagrams. }
\end{center}
\end{figure}

Many of these details can  be illustrated with a three-body system. Then, $G_0$ 
is the product of three full propagators $S$ and thus its gauged analogue 
generates three impulse-like diagrams
\begin{align}
	G_0^\mu=(S_1S_2S_3)^\mu=S_1^\mu S_2S_3+S_1S_2^\mu S_3+S_1S_2S_3^\mu=
	\chi^\mu_1 S_2S_3 + S_1\chi^\mu_2 S_3+ S_1 S_2\chi^\mu_3~.
\end{align}
Here $\chi^\mu=S^f\Gamma^\mu S^i$ and the superscripts $i,f$ denote that the 
propagators are evaluated before and after the \textit{momentum injection} from 
the external field. The interaction kernel $K$ must be decomposed into the sum 
of its two- and three-particle irreducible terms
\begin{align}
	K=\sum_{\textnormal{\tiny perm.}}K_{(\mathrm{2PI})} S^{-1}+K_{(\mathrm{3PI})}~.
\end{align}
Then, using \eqref{eq:definition_qphvertex_inverse} its gauged version is
\begin{align}\label{eq:gauged_kernel}
	K^\mu=\sum_{\textnormal{\tiny perm.}}K^\mu_{(\mathrm{2PI})} S^{-1}-\sum_{\textnormal{\tiny perm.}}K_{(\mathrm{2PI})} \chi^\mu+K^\mu_{(\mathrm{3PI})}~.
\end{align}
It is interesting to note how the gauging procedure has automatically introduced 
a relative sign between those terms in which the external field interacts with 
the spectator line and those in which it interacts with the irreducible kernel; 
this ensures the absence of the over counting of diagrams \cite{Kvinikhidze:1998xn,Kvinikhidze:1999xp}. 
The irreducible kernels themselves must be gauged once they are expressed in 
terms of the elementary degrees of freedom.
A diagrammatic representation of this equation is shown in Fig.~\ref{fig:FFeq_full}.

The application to two-body states is entirely analogous, with the 
simplification that $K^\mu$ in \eqref{eq:FFeq_compact} is directly the gauged 
two-particle irreducible kernel, $K^\mu_{(\mathrm{2PI})}$.
Also, the generalisation to the coupling of two external fields by 
\textit{gauging twice} has been presented in 
\cite{Eichmann:2011pv,Eichmann:2012mp}.

\section{Application to QCD}
In this section we apply the previously introduced formalism to QCD and derive
the quark self-energy, quark-gluon vertex and meson Bethe-Salpeter kernel
from the truncated 2PI and 3PI effective actions.

\begin{figure}[h]
\begin{center}
\includegraphics[scale=1.5]{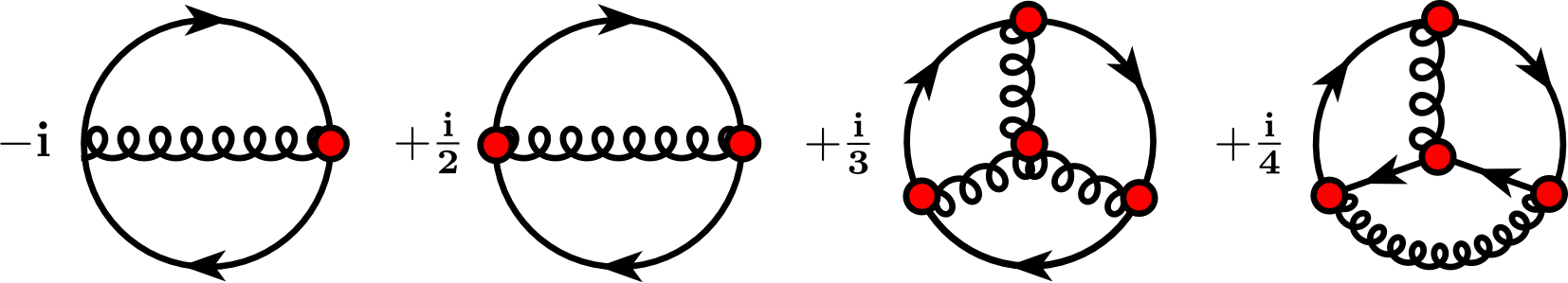}
\caption{\label{fig:action3PImeson}The terms of the 3PI effective action at 3-loop 
relevant to the quark, quark-gluon vertex and meson Bethe-Salpeter kernel.}
\end{center}
\end{figure}

\subsection{3PI Effective Action}

As an example, let us derive the quark self-energy, quark-gluon vertex, and 
meson Bethe-Salpeter kernel from the 3PI effective action at three-loop order. 
We start with the effective action, given in Fig.~\ref{fig:action3PImeson}, 
wherein we have kept only those terms relevant to the discussion at hand. Note 
also that at least a three-loop expansion is required in order to obtain, upon 
the inclusion of a \textit{baryonic vertex} in the effective action, a 
non-trivial baryon BSE kernel.

The quark self-energy is given by \eqref{eq:fermionselfenergy}, while solving 
the stationary condition $\delta\Gamma[\Psi,G,V]/\delta V^\mu_{ab}=0$ 
gives the quark-gluon vertex DSE
\begin{align}\label{eq:3PI_self_energy}
\Sigma = -i\frac{\delta \Gamma_2[\Psi,G,V]}{\delta G_{ab}} &= 
	\includegraphics[scale=1.3]{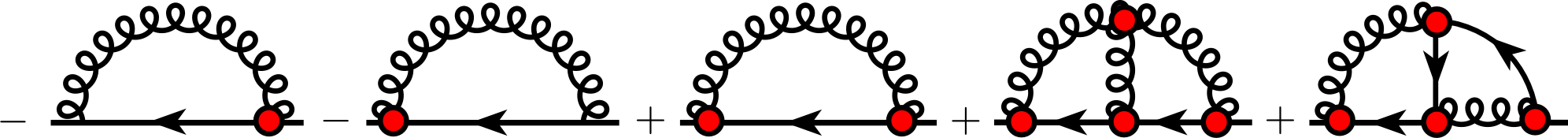} 
\nonumber\\ 
&=
	\includegraphics[scale=1.3]{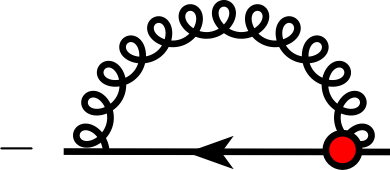}\;,
\end{align}
\begin{align}\label{eq:3PI_qgv}
	0 = \frac{\delta \Gamma_2[\Psi,G,V]}{\delta 
V^\mu_{ab}}\bigg|_{G=G^{(0)},V=V^{(0)}} \implies
	\begin{array}{c}
	\includegraphics[scale=1.3]{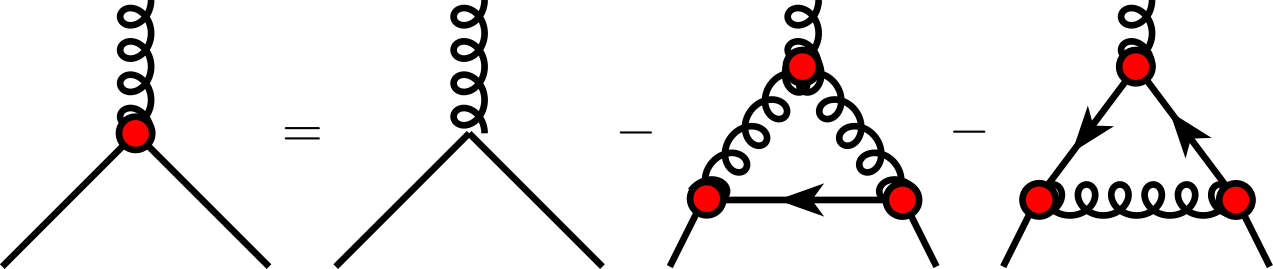} 
	\end{array}~.
\end{align}
In the second line of Eq.\eqref{eq:3PI_self_energy} we have used the 
vertex DSE \eqref{eq:3PI_qgv} to simplify the form of the self-energy 
contribution. Note that in doing so, the self-energy is no longer a function of 
$V$, but rather $V^{(0)}$ which now implicitly depends upon the quark 
propagators $G$. Thus, if the BSE kernel is obtained by a 
further functional derivative of the simplified self-energy, the functional 
dependence of $V^{(0)}$ on $G$ must be resolved.

Thus, we use the first line in Eq.~\eqref{eq:3PI_self_energy} to derive the 
Bethe-Salpeter kernel. Taking one further functional 
derivative of the quark self-energy with respect to the quark (which 
diagrammatically amounts to cutting one quark line) yields
\begin{align}\label{eq:bsekernel3PI}
    -K=\frac{\delta\Sigma}{\delta S} &=
    \begin{array}{c}
    \includegraphics[scale=1.3]{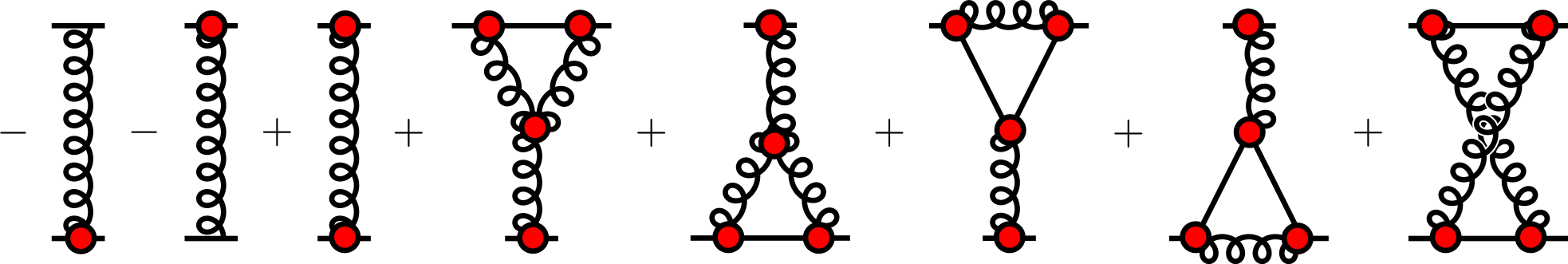} 
   	\end{array}\nonumber\\
   	&=
   	\begin{array}{c}
    \includegraphics[scale=1.3]{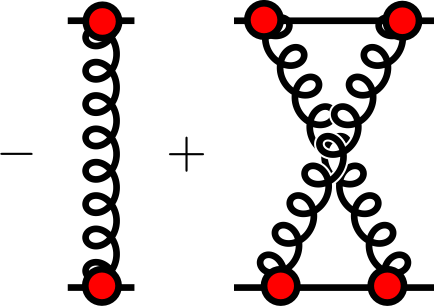} 
   	\end{array}\;,
\end{align}
where we can once again use the quark-gluon vertex DSE \eqref{eq:3PI_qgv} to 
simplify the kernel in the last step.

It is interesting to note here the appearance of a ladder exchange that features 
two dressed vertices. Additionally, at this order in the truncation the 
doubly-dressed gluon exchange must necessarily be accompanied by a crossed-ladder 
exchange in the BSE kernel in order to preserve chiral symmetry and any other 
global symmetries of the system.

\subsection{2PI Effective Action}
It is enlightening to compare the 3PI effective action at 3-loop to the 2PI
effective action at the same order. We can read this off from 
Fig.~\ref{fig:action3PImeson} by replacing the dressed vertices with bare ones.
Then, the quark self-energy is 
\begin{align}\label{eq:2PI_self_energy}
\Sigma = -i\frac{\delta \Gamma_2[\Psi,G]}{\delta G_{ab}} &= 
	\includegraphics[scale=1.3]{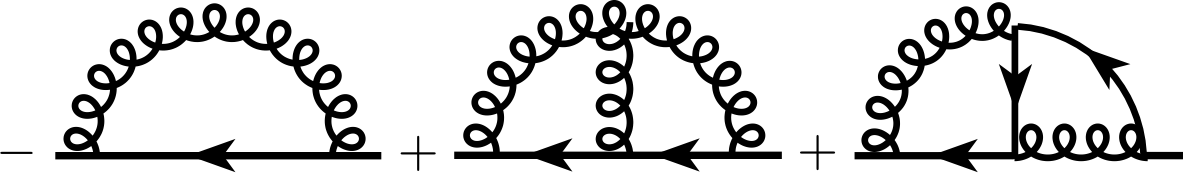} 
\nonumber\\ 
&=
	\includegraphics[scale=1.3]{figures/action_3PI_3loop_quark_2}\;,
\end{align}
from which the quark-gluon vertex can be inferred
\begin{align}\label{eq:2PI_qgv}
 \implies
	\begin{array}{c}
	\includegraphics[scale=1.3]{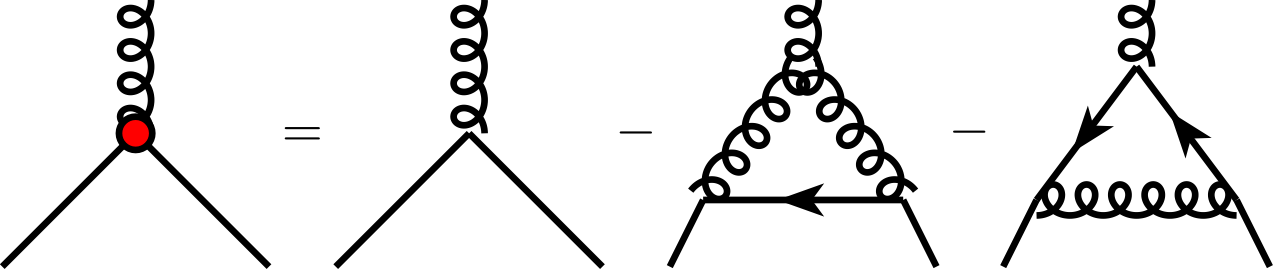} 
	\end{array}~.
\end{align}
In a similar fashion to the above, we can write down the corresponding 
Bethe-Salpeter kernel by taking one further functional 
derivative of the quark self-energy with respect to the quark (resolving
the quark dependence of the vertex if needs be) to find
\begin{align}\label{eq:bsekernel2PI}
    -K=\frac{\delta\Sigma}{\delta S} &=
    \begin{array}{c}
    \includegraphics[scale=1.3]{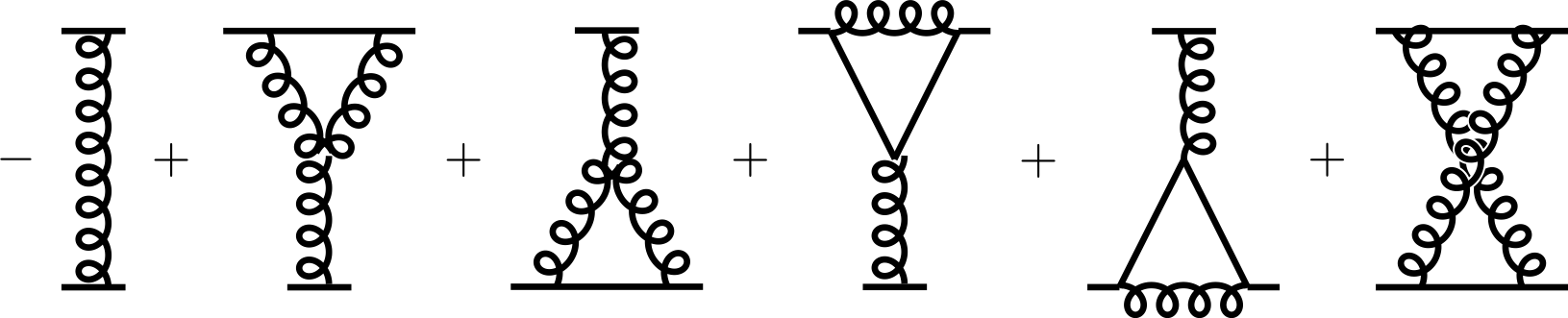} 
   	\end{array}\nonumber\\
   	&=
   	\begin{array}{c}
    \includegraphics[scale=1.3]{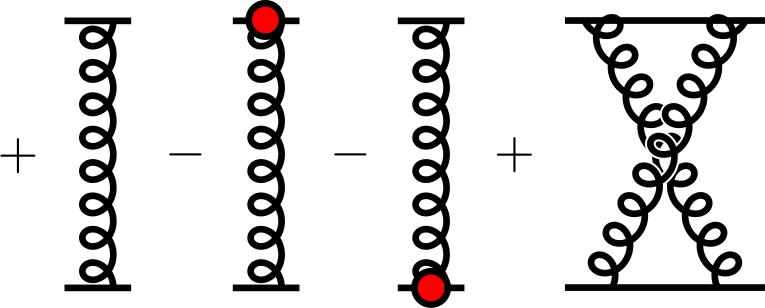} 
   	\end{array}\;.
\end{align}
This truncation is essentially the one employed in Refs.~\cite{Williams:2014iea,Vujinovic:2014ioa};
the application to baryons is reported in Ref.~\cite{Sanchis-Alepuz:2015qra}.

Notice here that \eqref{eq:bsekernel2PI} is structurally quite different from
\eqref{eq:bsekernel3PI} in that the ladder exchanges always contain one 
perturbative vertex; this could be remedied by including, for example, 4-loop 
terms in the 2PI effective action. 

All of this can be compared to the rainbow-ladder truncation which follows from
the 2PI effective action at two-loop order. Then only the two-loop terms (with 
vertices bare) of Fig.~\ref{fig:action3PImeson} are required, yielding
\begin{align}\label{eq:dseandbseRL}
\Sigma &= \includegraphics[scale=1.3]{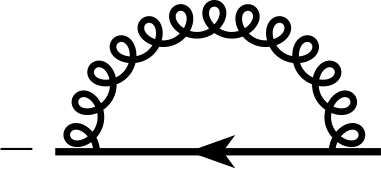}\;\;,
\qquad\qquad-K=	
   	\begin{array}{c}
    \includegraphics[scale=1.3]{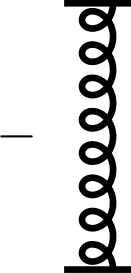} 
   	\end{array}\;.
\end{align}
Note that only this simplistic truncation lacks the flavour dependence
that the two-body kernel necessarily features (due to implicit and explicit
dependence on the quark propagator). To make such a truncation viable, 
the bare vertices are renormalisation-group improved (i.e. dressed such that 
perturbative anomalous dimensions are recovered). This accounts for the lack of 
interaction strength provided by a single gluon-exchange, and is important for 
both phenomenology and chiral dynamics.

\section{Some technical remarks}\label{sec:technicalitites}
We comment in this last section upon some technical aspects that must be taken 
into account when attempting to solve, in practice, the DSE/BSE system.
\subsection{Covariant decomposition of amplitudes}
The quantum numbers, such a spin, parity, etc. of the bound state to be studied 
can be enforced by restricting the tensor structure of the Bethe-Salpeter 
amplitudes to have the correct symmetries for these quantum numbers.

For the description of a two-fermion bound-state, we need to provide a covariant 
decomposition $\psi^\mathcal{I}_{\alpha\beta}$. The spinor indices of the two 
fermions are $\alpha,\beta$, while for  total spin $J$, $\mathcal{I}$ is a 
product of $J$ Lorentz indices. For spin $J=0$ it is well-known that a suitable 
representation is~\cite{Joos:1962qq,Weinberg:1964cn}
\begin{align}\label{eqn:diracmatrices}
\mathds{1},\; \gamma^\mu,\; \gamma^{[\mu}\gamma^{\nu]},\;\gamma^{[\mu}\gamma^{\nu}\gamma^{\rho]},\;\gamma^{[\mu}\gamma^{\nu}\gamma^{\rho}\gamma^{\sigma]}\;,
\end{align}
that is often simplified through the introduction of $\gamma_5$. Saturating these with the quark relative momentum $k^\mu$ and imposing 
positive parity yields $D_i=\left\{ \mathds{1}, \slashed{k}\right\}$. Then, the 
general decomposition for a state of zero spin is
\begin{align}
	\Gamma(k,P)&=
	\left(\begin{array}{c}\ONE \\ \gamma_5\end{array}\right) D_i\; \Lambda_{\pm}\;.
\end{align}
Here, the first term selects the overall parity of the state and 
$\Lambda_\pm = \left( \mathds{1} \pm \hat{\slashed{P}}\right)/2$ is a 
positive/negative energy projector that introduces the (normalised) total 
bound-state momentum $P$.  
To introduce total angular momentum $J$, we couple the spin-zero state $\Gamma$ 
to the two possible angular momentum tensors $Q^{\mu_1\ldots\mu_J}$ and 
$T^{\mu_1\ldots\mu_J}$
\begin{align}
	\Gamma^{\mu_1\ldots\mu_J}(k,P) = 
	\left(\begin{array}{c}Q^{\mu_1\ldots\mu_J} \\ T^{\mu_1\ldots\mu_J}\end{array}\right)
	\Gamma(k,P)\;.
\end{align}
These tensors are given by the traceless part of the symmetrised J-fold tensor 
products of a transversal projector transforming like a 
vector~\cite{Zemach:1968zz}.

It is convenient to introduce the transverse projector 
$T^{\mu\nu}_P = \delta^{\mu\nu}-P^\mu P^\nu/P^2$, and to denote its application 
using subscripts as follows:
$k^\mu_T =  T^{\mu\nu}_P k^\nu$; 
$\gamma^\mu_t = T^{\mu\alpha}_k T^{\alpha\nu}_P\gamma^\nu$. If we define the 
symmetrised $J$-fold tensor products
\begin{align}
	\tilde{Q}^{\mu_1\ldots\mu_J} = k^{\{\mu_1}_T \cdots k^{\mu_J\}}_T\;,\;\;\; 
	\tilde{T}^{\mu_1\ldots\mu_J}=\gamma^{\{\mu_1}_t k^{\mu_2}_T \cdots k^{\mu_J\}}_T\;,
\end{align}
then the angular momentum tensors $Q^{\mu_1\ldots\mu_J}$ and 
$T^{\mu_1\ldots\mu_J}$ are the traceless part 
thereof~\cite{LlewellynSmith:1969az,Krassnigg:2010mh,Fischer:2014xha}. 

To describe a three-fermion bound-state we need to provide a covariant 
decomposition for $\psi^\mathcal{I}_{\alpha\beta\gamma}$. The spinor indices 
$\alpha,\beta,\gamma$ correspond to the three fermionic legs, while for total 
spin-$k+\nicefrac{1}{2}$, $\mathcal{I}$ is composed of one spinor and $k$ 
Lorentz indices.

Let us begin with a spin-$\nicefrac{1}{2}$ baryon where $\mathcal{I}=\delta$ 
carries the incoming baryon spin index. Saturating the matrices 
\eqref{eqn:diracmatrices} with the two relative quark momenta $k^\mu$ and 
$q^\mu$ and selecting positive parity gives 
$D_i=\left\{ \mathds{1}, \slashed{k}_T, \slashed{q}_t, \slashed{k}_T\slashed{q}_t\right\}$. 
Then
\begin{align}
	\psi_{\alpha\beta;\gamma}^{\phantom{\alpha\beta;\gamma}\delta}   (k,q,P)=
	\left(\begin{array}{ccc}
	\mathds{1} &           \otimes & \mathds{1}  \\ 
	\gamma_5 &             \otimes & \gamma_5    \\ 
	\gamma^\rho_T&          \otimes & \gamma^\rho_T\\ 
	\gamma^\rho_T\gamma_5 & \otimes & \gamma^\rho_T\gamma_5 
	\end{array}\right)
	\left(\begin{array}{ccc}
	D_i & \otimes & D_j
	\end{array}\right)
	\left(\begin{array}{ccc}
	\Lambda_\pm\gamma_5 C & \otimes & \Lambda_+
	\end{array}\right)\;.
\end{align}
The left tensor product denotes the outgoing quark legs with indices $\alpha$, 
$\beta$ and hence warrants the inclusion of $\gamma^5C$, with 
$C=\gamma^4\gamma^2$ the charge conjugation matrix. The right tensor product 
describes the outgoing quark leg, index $\gamma$, and the incoming baryon 
spin-index $\delta$; the $\Lambda_+$ here selects the positive energy baryon. 
Not all elements are linearly independent; it can be checked that a linearly 
independent subspace of 64 
elements can be constructed \cite{Eichmann:2009qa,Eichmann:2009en}.

The generalisation to a state of spin-$k+\nicefrac{1}{2}$ (with $k$ integer) is 
obtained by extension of this basis
\begin{align}
	\psi_{\alpha\beta;\gamma}^{\phantom{\alpha\beta;\gamma}\delta\nu_1\ldots\nu_k}(k,q,P)&=
	\left(\begin{array}{ccc}
 	M^{\mu_1\ldots\mu_k} & \otimes & \mathds{P}^{\mu_1\ldots\mu_k\nu_1\ldots\nu_k}
	\end{array}\right)
	\psi_{\alpha\beta;\gamma}^{\phantom{\alpha\beta;\gamma}\delta}(k,q,P)\;, \\
	M^{\mu_1\ldots\mu_k} &= \left\{\begin{array}{c}
	\gamma^{\mu_1}_T \cdots \gamma^{\mu_k}_T \gamma_5\\
	\gamma^{\mu_1}_T \cdots \hat{k}^{\mu_k}_T \gamma_5\\
	\hat{k}^{\mu_1}_T \cdots \hat{q}^{\mu_k}_t \gamma_5\\
	\ldots\\
	\end{array}\right.\;,
\end{align}
with $M^{\mu_1\ldots\mu_k}$ representing all combinations of $k$ products of $\gamma^{\mu_i}_T$, $k^{\mu_i}_T$ and $q^{\mu_i}_t$.
Here, $\mathds{P}^{\mu_1\ldots\mu_k\nu_1\ldots\nu_k}$ is the generalised Rarita-Schwinger 
projector. A linearly independent subspace spans $64(k+1)$ elements. As an example, we give the Rarita-Schwinger projector for 
spin-$\nicefrac{3}{2}$
\begin{align}
	\mathds{P}^{\mu_1\nu_1} = T^{\mu_1\nu_1}_P - \frac{1}{3}\gamma^{\mu_1}_T\gamma^{\nu_1}_T\;,
\end{align}
where note that we omitted the here redundant positive energy projector 
$\Lambda_+$. In this case $128$ linearly independent basis-elements can be 
constructed \cite{SanchisAlepuz:2011jn}.

\subsection{Euclidean spacetime and quarks in the complex plane}

\begin{figure}[t!]
\begin{center}
\includegraphics[scale=0.5]{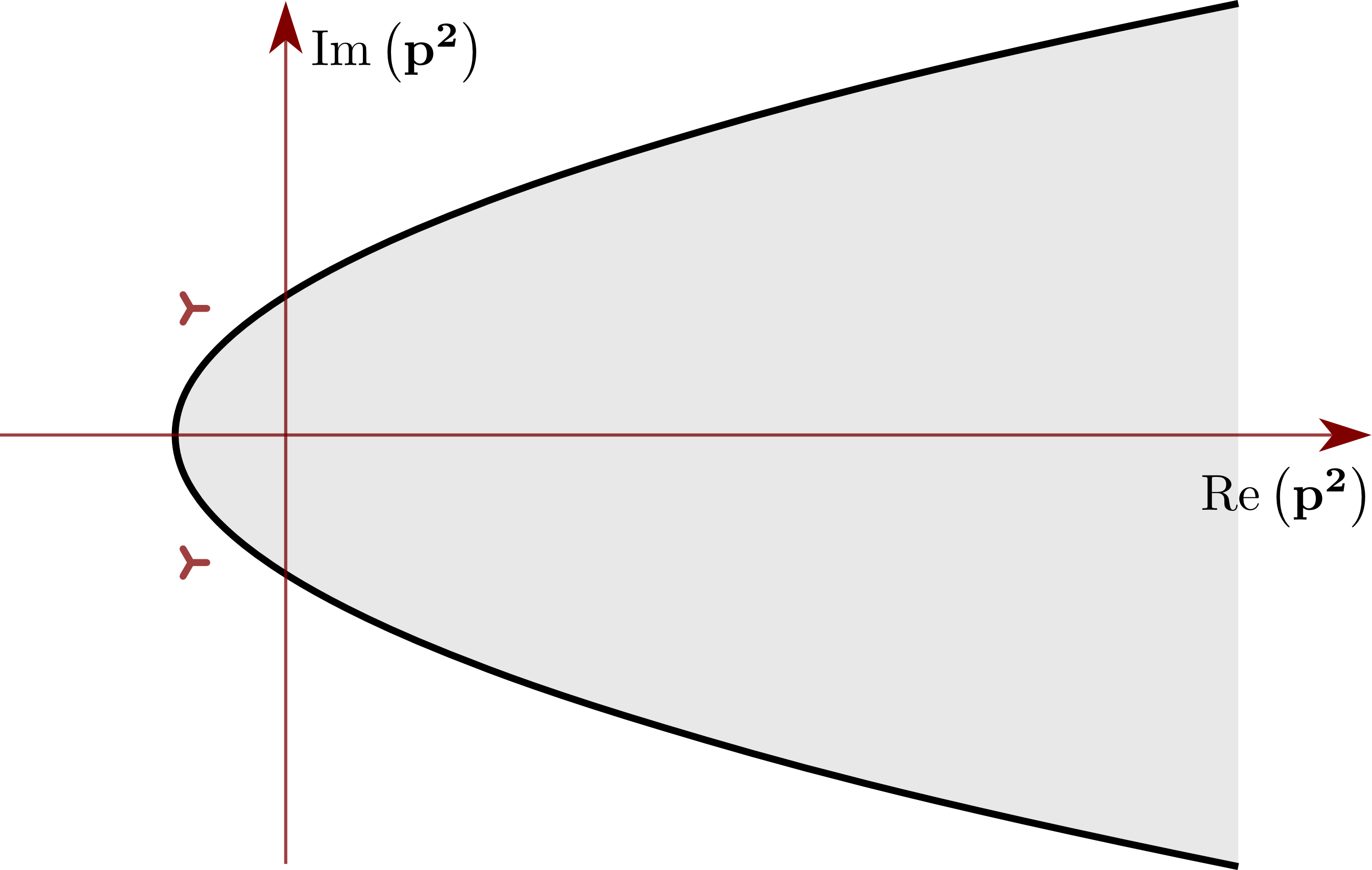}
\caption{\label{fig:parabola}A sketch of the bounded parabolic region of the 
complex plane probed by the quark propagators in the Bethe-Salpeter equation. 
Crosses symbolise the appearance of complex conjugate poles in the timelike 
complex region.}
\end{center}
\end{figure}

Calculations using DSEs and BSEs are mainly performed in Euclidean momentum 
space (for studies using Minkowski spacetime see 
\cite{Sauli:2014uxa,Biernat:2014eya} and references therein). In particular, 
this means that for a bound state of mass $M$, the total momentum $P$ in the 
rest frame is 
\begin{align}
    P=\left( 0,0,0,i\,M \right)~.
\end{align}
Let us focus, for simplicity, on the case of a mesonic bound state. If the 
relative momentum between the constituent quarks is $k$, the momentum of the 
constituents can be written as
\begin{align}\label{eq:quark_momenta}
    p_{\pm}=k\pm \frac{P}{2}~.
\end{align}
The propagators for these constituents are functions of $p_{\pm}^2$. Using 
\eqref{eq:quark_momenta} and assuming that the relative momentum $k$ is real, 
one can see that
\begin{align}
    p_{\pm}^2=\left( t\pm i~M\right)^2
\end{align}
for some real and positive parameter $t$. This is the parametric 
equation for a parabola. Therefore, the quark dressing functions need to be 
known in a parabolic region of the complex plane (see Fig.~\ref{fig:parabola}).
It is a general feature of the analytic structure of the quark propagators in 
the complex plane to have complex conjugate poles in the timelike complex 
region. These poles pose a limitation on the maximum mass of the bound state 
that can be calculated as well as the spacelike region for which form factors 
can be studied (see, e.g. \cite{Eichmann:2011aa}). The possibility of 
parametrising the quark propagators by analytic functions that allow better 
control over its singularities has been explored, \emph{e.g.} in 
\cite{Bhagwat:2002tx,Dorkin:2013rsa,Mezrag:2014jka}.

\section{Conclusions}
In these proceedings we have collected the key aspects for
a systematic study of hadronic properties such as masses and form-factors
using the combination of Dyson-Schwinger and Bethe-Salpeter equations.
The use of nPI effective action techniques proves to be a powerful resource 
in this respect.

It should be clear from the presentation that the now ubiquitous 
rainbow-ladder truncation is the first term in a systematic expansion
of the Bethe-Salpeter kernel using the effective action; 
in particular it appears when a renormalisation-group improved (RGI) 2PI effective action at 
two-loops is used. Present investigations~\cite{Sanchis-Alepuz:2015qra} are aimed at 
extending the calculation of meson and baryon spectra using kernels derived from
a three-loop truncation of the effective action.

%
%
%
%
\ack
HSA would like to thank the organisers of ``Discrete 2014'' for their 
hospitality.
This work has been supported by an Erwin Schr\"odinger fellowship J3392-N20 
from the Austrian Science Fund (FWF), by the Helmholtz International Center for 
FAIR within the LOEWE program of the State of Hesse, and by the DFG 
collaborative research center TR 16.

%
%
%
%

\section*{References}
\bibliographystyle{iopart-num}
\bibliography{Sanchis-Alepuz_Helios_DISCRETE2014}
\end{document}